\documentclass[rnote]{aa} 
\usepackage{graphicx}
\usepackage{txfonts}
\usepackage{longtable}

\begin{document}
   \title{Low-amplitude and long-period radial velocity variations in giants HD 3574, 63 Cygni, and ~HD~216946\thanks{Based on observations made with the BOES instrument on the 1.8 m telescope at Bohyunsan Optical Astronomy Observatory in Korea.}}
   \author{B.-C. Lee\inst{1},
          I. Han\inst{1},
          M.-G. Park\inst{2},
          A. P. Hatzes\inst{3},
          \and
          K.-M. Kim \inst{1}
          }

   \institute{Korea Astronomy and Space Science Institute, 776,
		Daedeokdae-Ro, Youseong-Gu, Daejeon 305-348, Korea\\
	      \email{[bclee;iwhan;kmkim]@kasi.re.kr}
	    \and
	     Department of Astronomy and Atmospheric Sciences,
	     Kyungpook National University, Daegu 702-701, Korea\\
	      \email{mgp@knu.ac.kr}
        \and
        Th{\"u}ringer Landessternwarte Tautenburg (TLS), Sternwarte 5, 07778 Tautenburg, Germany\\
	      \email{artie@tls-tautenburg.de}
             }

   \date{Received 9 May 2013 / Accepted 19 May 2014}


  \abstract
   {}
   {We study the low-amplitude and long-period variations in evolved stars  using precise radial velocity measurements.
   }
   {The high-resolution, fiber-fed Bohyunsan Observatory Echelle Spectrograph (BOES) was used from September 2004 to May 2014 as part of the  exoplanet search program at the Bohyunsan Optical Astronomy Observatory (BOAO).
   }
   {We report the detection of low-amplitude and long-period orbital radial velocity variations in three evolved stars, HD~3574, 63~Cyg, and~HD 216946. They have periods of 1061, 982, and 1382 days and semi-amplitudes of 376, 742, and 699 m s$^{-1}$, respectively.
   }
   {}


   \authorrunning{B.-C. Lee et al.} 
   \titlerunning{Low-amplitude and long-period radial velocity variations in giants HD 3574, 63 Cygni, and HD 216946}
   \maketitle
%

\section{Introduction}
Over the past two decades, the development of high-resolution spectroscopies and stable wavelength references enabled the precise measurement of radial velocities (RV) of stars. This opened up the field of exoplanet and brown dwarf research.
While the exoplanets have been detected with a frequency of $\sim$ 5\% for dwarf stars and $\sim$ 10\% for giant stars, brown dwarf companions are estimated to be around fewer than 1\% of Sun-like stars within 3 AU (Kraus et al. 2008).
This dramatic dichotomy indicates that perhaps the formation of brown dwarfs and stellar companions involves very different mechanisms. Low-mass stars and brown dwarfs can form through the fragmentation of dense filaments and disks, possibly followed by early ejection from these dense environments, which truncates their growth in mass (Bonnell et al. 2007). The formation of brown dwarfs is still under dispute  and it is unclear whether brown dwarfs could form via planetary formation processes. Finding more brown dwarfs can provide a key to star and planet formation because they represent the extreme mass limit of both processes.

The primary aim of this work is to find periodic RV variations caused by substellar-mass companions.
In this paper, we present precise RV measurements for three evolved stars. In Sect. 2, we briefly describe the observations and data reduction. The stellar properties including photometric variations are presented in Sect. 3. We examine the RV measurements and orbital solutions in Sect. 4. Finally, in Sect. 5, we discuss our results.


\begin{table*}
\begin{center}
\caption[]{Stellar parameters.}
\label{tab1}
\begin{tabular}{lcccc}
\hline
\hline
    Parameters              & HD 3574     & 63 Cyg &  HD 216946 &  Reference     \\

\hline
    Spectral type             & K5 III  & K4 Ib-II & M0 Iab var & \emph{HIPPARCOS} (ESA 1997)   \\
    $\textit{$m_{v}$}$ [mag]  & 5.45    & 4.56     & 4.99       & \emph{HIPPARCOS} (ESA 1997)  \\
    $\textit{B-V}$ [mag]      & 1.644 $\pm$ 0.015 & 1.569 $\pm$ 0.012 & 1.778 $\pm$ 0.013 & van Leeuwen (2007) \\
    $\emph{d}$ [pc]           & --  & 314.46 $\pm$ 22.95 & 480.82 $\pm$ 61.06 & Anderson \& Francis (2012) \\
                  & 641.0 $^{+178.7}_{-114.7}$ & 316.5 $^{+24.8}_{-21.5}$ & 490.2 $^{+71.6}_{-55.4}$ & Derived \\
    RV [km s$^{-1}$]     & -- 6.6 $\pm$ 0.8 & -- 26.32 $\pm$ 0.23 & -- 9.5 $\pm$ 0.7 & Kharchenko et al. (2007) \\

    $\pi$ [mas]               & 1.56 $\pm$ 0.34  & 3.16 $\pm$ 0.23 & 2.04 $\pm$ 0.26 & van Leeuwen (2007) \\
    $T_{\mathrm{eff}}$ [K]    & 4420   & 4080 &    3650        & Wright et al. (2003) \\
                & 4399 $^{+357}_{-530}$ & 4170 $^{+226}_{-399}$ & 4680 $^{+593}_{-495}$ & Ammons et al. (2006) \\
                              & 3830     & --   & -- & S{\'a}nchez-Bl{\'a}zquez et al.(2006)  \\
                              & --       & --        & 3800        & Massey et al. (2009)     \\
                              & --           &  4150  & --        & Lafrasse et al. (2010)    \\
                              & --       &  --    & 4000 $\pm$ 200  & Soubiran et al. (2010)  \\
                              & 4048 $\pm$ 28   &  --    &   --  & Prugniel et al. (2011)  \\
    log $\it g$               & --              & 1.9            & --               & Lafrasse et al. (2010)  \\
                              & 1.44            & --   & --  & S{\'a}nchez-Bl{\'a}zquez et al. (2006)  \\
                              & --              & --             & 0.5 $\pm$ 0.3    & Soubiran et al. (2010) \\
                              & 1.13 $\pm$ 0.14 & --             & --               & Prugniel et al. (2011)  \\
    $\mathrm{[Fe/H]}$         & -- 0.01 & --  & --  & S{\'a}nchez-Bl{\'a}zquez et al. (2006)     \\
                              & --      & --  & -- 0.07 $\pm$ 0.13   & Soubiran et al. (2010)     \\
                              & 0.10 $\pm$ 0.05 & --     & --       & Prugniel et al. (2011)     \\
                              & 0.08 $\pm$ 0.06 & --     & -- 0.02  & Anderson \& Francis (2012) \\
                              & --   &  --    & -- 0.03   & Cesetti et al. (2013) \\
    $\textit{$R_{\star}$}$ [$R_{\odot}$] & --      & 35        & 350      & Pasinetti-Fracassini et al. (2001) \\
    $\textit{$M_{\star}$}$ [$M_{\odot}$] & 7.7 $\pm$ 0.4  & --       & 6.8 $\pm$ 1.0  & Tetzlaff et al. (2011) \\

\hline

\end{tabular}
\end{center}
\end{table*}
%


\section{Observations and reduction}
%

We conducted a precise RV survey of about 300 evolved stars using the fiber-fed high-resolution ($\emph{R}$ = 90 000) Bohyunsan Observatory Echelle Spectrograph (BOES; Kim et al. 2007), which covers a wavelength region from 3500 to 10~500\,${\AA}$. To provide precise RV measurements, the spectrograph is equipped with an  iodine absorption (I$_{2}$) cell that produces reference absorption lines in the  region of 4900$-$6000 {\AA}. Observations of an RV standard star, $\tau$ Cet,  show long-term rms scatter of $\sim$ 7 m s$^{-1}$ (Lee et al. 2013).

From September 2004 to May 2014, we obtained 41, 34, and 32 spectra each for HD 3574, 63 Cyg, and HD 216946, respectively. The  signal-to-noise ratio (S/N) at the  I$_{2}$ region was about 200 with a typical exposure time of between 6 and 15 minutes. The basic reduction of spectra was performed with the IRAF (Tody 1986) software package and the code DECH (Galazutdinov 1992). RV measurements were computed  using the RVI2CELL (Han et al. 2007) based on a method by Butler et al. (1996) and Valenti et al. (1995). The resulting RV measurements are listed in Tables~\ref{tab2}, ~\ref{tab3}, and ~\ref{tab4}.
%


\section{Stellar characteristics}
\subsection{Fundamental parameters}
The basic stellar  parameters were taken from the \emph{HIPPARCOS} satellite catalog (ESA 1997). HD 3574 (= HR~164 = HIP~3083) is a K5 III, a visual magnitude of 5.45 with  a color index of 1.644 mag. The star 63 Cyg (= HD~201251 = HR~8089 = HIP~104194) is a K4 Ib-II, a visual magnitude of 4.56, and it has a color index of 1.569 mag. Lastly, HD~216946 (= HR~8726 = HIP~113288 = V424~Lac) is classified in SIMBAD as a pulsating variable star and is an M0 Iab with a visual magnitude of 4.99 and a color index of 1.778 mag. Astrometric parallaxes, $\pi$, we taken from the improved \emph{HIPPARCOS} values  by van Leeuwen (2007). The corresponding values are 1.56 $\pm$ 0.34 (HD~3574), 3.16 $\pm$ 0.23 (63~Cyg), and 2.04 $\pm$ 0.26 mas (HD 216946). These  imply distances (\emph{d}) of 641.0 $^{+178.7}_{-114.7}$ (HD 3574), 316.5 $^{+24.8}_{-21.5}$ (63~Cyg), and 490.2 $^{+71.6}_{-55.4}$ pc (HD~216946), respectively. For 63~Cyg and HD~216946 these are roughly similar to the values of Anderson \& Francis (2012). Stellar radii and masses were taken from Pasinetti-Fracassini et al. (2001) and Tetzlaff et al. (2011). The estimated parameters yielded $R_{\star}$ = 35 (63 Cyg) and 350 $R_{\odot}$ (HD~216946) and $M_{\star}$ = 7.7 $\pm$ 0.4 (HD 3574) and 6.8 $\pm$ 1.0 $M_{\odot}$ (HD~216946). The basic stellar parameters are summarized in Table~\ref{tab1}.

\subsection{Photometric variations}
\begin{figure}
   \centering
   \includegraphics[width=8cm]{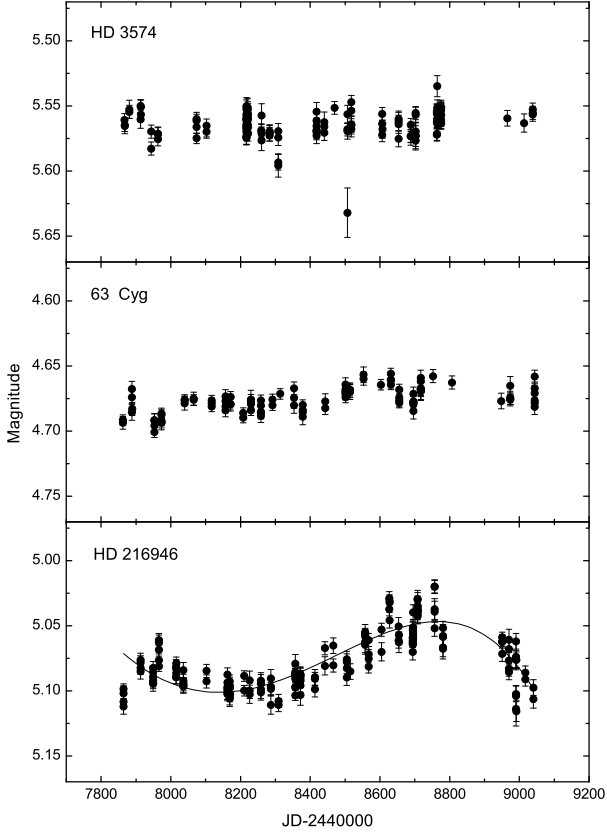}
      \caption{\emph{HIPPARCOS} photometric variations for HD 3574, 63 Cyg, and HD 216946 (\emph{top} to \emph{bottom panel}), respectively. HD 216946 shows low-amplitude and long-term photometric variations with a period of 1100 days (curve).
        }
        \label{Hip}
   \end{figure}

%
We examined the  \emph{HIPPARCOS} photometry data obtained from December 1989 to February 1993 to search for such variations. There is a total of  123, 99, and 173 \emph{HIPPARCOS} measurements for HD 3574, 63 Cyg, and HD 216946, respectively. The former two stars show photometric stability down to an rms scatter of 0.011 mag for HD 3574 and 0.010 mag for 63 Cyg mag. This corresponds to 0.20 and 0.22\% variations over the time span of the observations, but we note that the \emph{HIPPARCOS} measurements where not contemporaneous with our spectral observations. HD~216946 shows brightness variations of 0.60\%, corresponding to an rms scatter of 0.030 mag, which was calculated  by excluding two measurements of 5.6 (JD 2448990.27697) and 10.2 (JD 2449017.06776) magnitude that were clear outliers. Figure~\ref{Hip} shows the time series of the \emph{HIPPARCOS} data. Interestingly,  HD 216946 reveals periodic variations of $\sim$ 1100 days with an amplitude of 0.05 mag.

Photometric observations for HD 216946 by Messina (2007) using the 80 cm automated photometric telescope (APT-80)\footnote{located at the M.G. Fracastoro station of INAF-Catania Astrophysical Observatory on Mt. Etna in Italy} were obtained from August 1993 to October 2001 for a total of 232 nights. The data show variations of 1601 days that  increased towards shorter wavelengths, from 0.08 mag for V band, 0.11 mag for B band, to 0.33 mag for U band.

%

\section{Radial velocity and orbital solutions}
%

%
   \begin{figure}
   \centering
   \includegraphics[width=8cm]{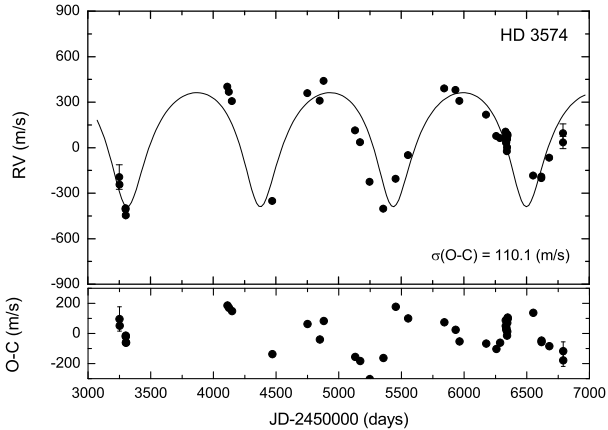}
      \caption{Variations of the RV curve (\emph{top panel}) and rms scatter of the residual (\emph{bottom panel}) for HD 3574 from September 2004 to May 2014. The solid line is the orbital solution with a period of 1061.2 days and an eccentricity of 0.32.
              }
         \label{rv1}
   \end{figure}
%
%
   \begin{figure}
   \centering
   \includegraphics[width=8cm]{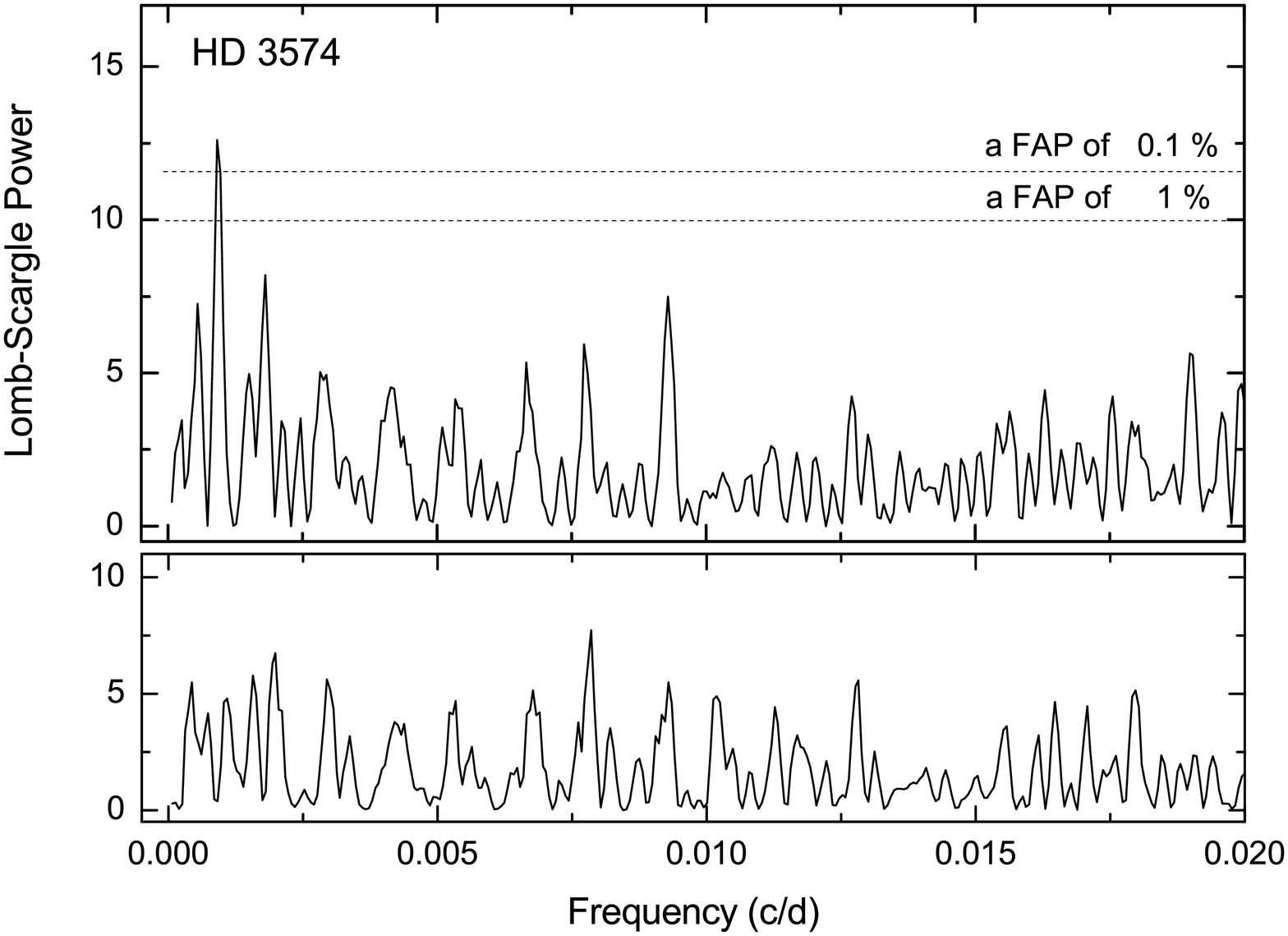}
      \caption{Lomb-Scargle periodogram of the RV measurements for HD 3574. The periodogram shows significant power at a period of 1061.2 days (\emph{top panel}) and after subtracting the main frequency variations (\emph{bottom panel}). The horizontal lines indicate FAP thresholds of 0.1 and 1\%.
      }
         \label{power1}
   \end{figure}
%
%
\begin{table}
\begin{center}
\caption{RV measurements for HD 3574 from September 2004 to May 2014 using the BOES.}
\label{tab2}
\begin{tabular}{rrrrrr}
\hline\hline

 JD         & $\Delta$RV  & $\pm \sigma$ &        JD & $\Delta$RV  & $\pm \sigma$  \\
 $-$2 450 000 & m\,s$^{-1}$ &  m\,s$^{-1}$ & $-$2 450 000  & m\,s$^{-1}$ &  m\,s$^{-1}$  \\
\hline
3251.1363 &   $-$193.0  &   80.8 &  5962.9131  &    308.4  &    9.9   \\
3253.2505 &   $-$243.6  &   10.2 &  6176.2766  &    217.7  &    8.6   \\
3301.1988 &   $-$406.3  &    8.6 &  6257.9621  &     77.2  &    8.1   \\
3301.2066 &   $-$398.8  &    7.4 &  6288.0551  &     63.0  &    9.2   \\
3303.1923 &   $-$446.8  &    9.0 &  6331.9416  &     72.0  &    9.4   \\
3303.1996 &   $-$445.6  &    8.8 &  6332.9335  &    105.8  &    8.8   \\
4111.9405 &    402.5  &    9.3 &  6334.9348  &     55.4  &   11.0   \\
4124.9567 &    367.5  &    9.7 &  6336.9164  &     31.3  &    8.8   \\
4147.9271 &    306.7  &   10.0 &  6337.9603  &     36.6  &   11.8   \\
4469.8840 &   $-$351.3  &   10.0 &  6342.9208  &    $-$22.7  &    8.2   \\
4750.9820 &    359.5  &    9.9 &  6343.9271  &      1.7  &   10.0   \\
4848.9819 &    309.6  &    9.3 &  6344.9290  &     56.7  &    9.3   \\
4879.9255 &    440.2  &    8.3 &  6346.9246  &     87.0  &   10.2   \\
5131.0493 &    114.6  &    9.7 &  6346.9313  &     78.4  &    9.5   \\
5171.9347 &     36.7  &    9.7 &  6552.0132  &   $-$184.0  &    9.9   \\
5247.9440 &   $-$225.1  &    9.5 &  6616.9237  &   $-$200.9  &    8.9   \\
5356.2589 &   $-$401.6  &    9.2 &  6616.9278  &   $-$190.1  &    9.1   \\
5455.3551 &   $-$204.5  &    9.3 &  6679.0684  &    $-$66.3  &   11.0   \\
5554.0040 &   $-$48.8   &   9.3  & 6790.3053   &    34.3   &  40.7    \\
5841.9482 &   390.7   &  10.6  & 6790.3114   &    95.8   &  60.5    \\
5932.9486 &   380.2   &   9.9  &               &           &          \\
\hline

\end{tabular}
\end{center}
\end{table}
%

%
   \begin{figure}
   \centering
   \includegraphics[width=8cm]{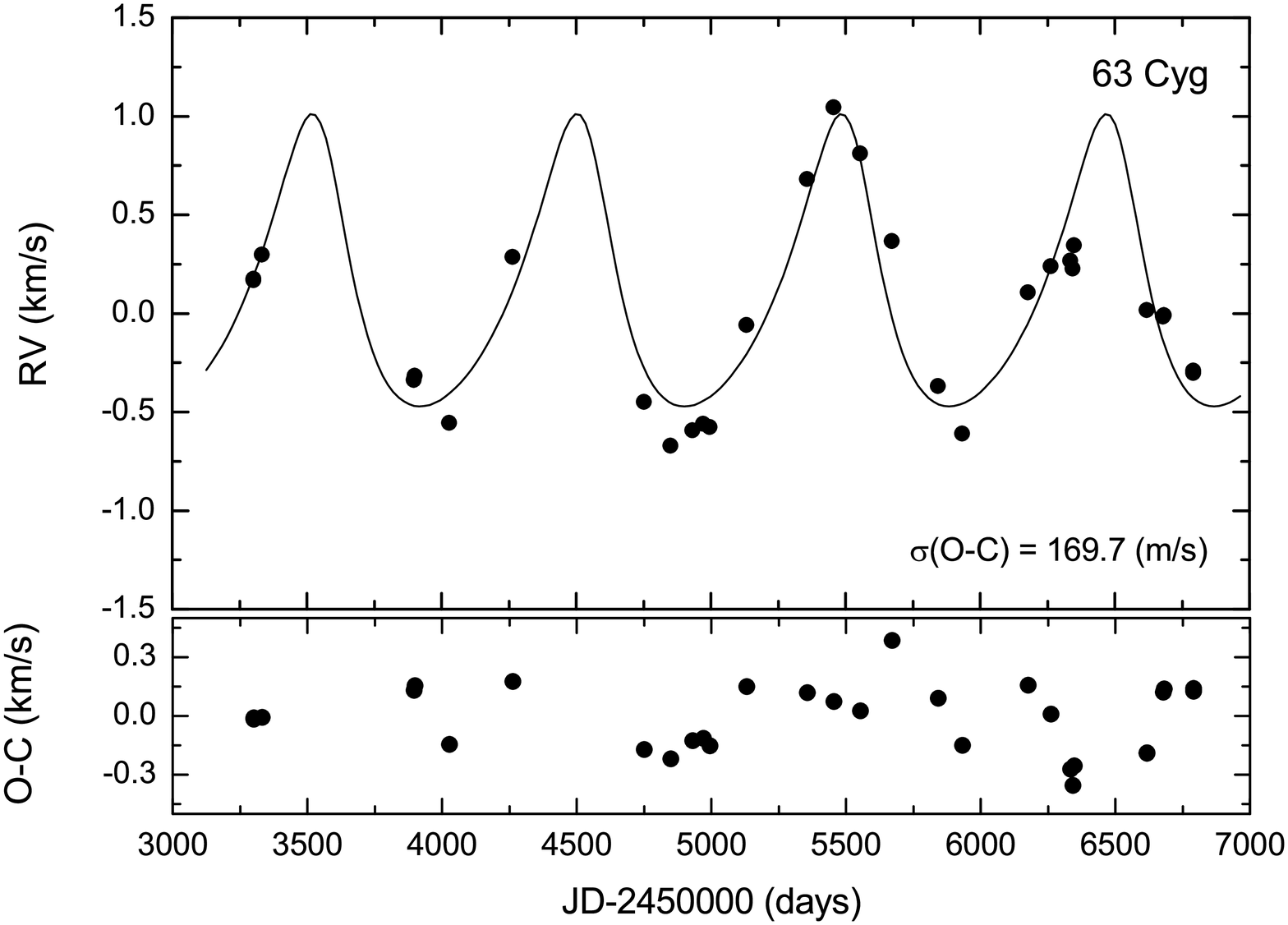}
      \caption{Variations of the RV curve (\emph{top panel}) and rms scatter of the residual (\emph{bottom panel}) for 63 Cyg from October 2004 to May 2014. The solid line is the orbital solution with a period of 982.8 days and an eccentricity of 0.29.
              }
         \label{rv2}
   \end{figure}
%
%
   \begin{figure}
   \centering
   \includegraphics[width=8cm]{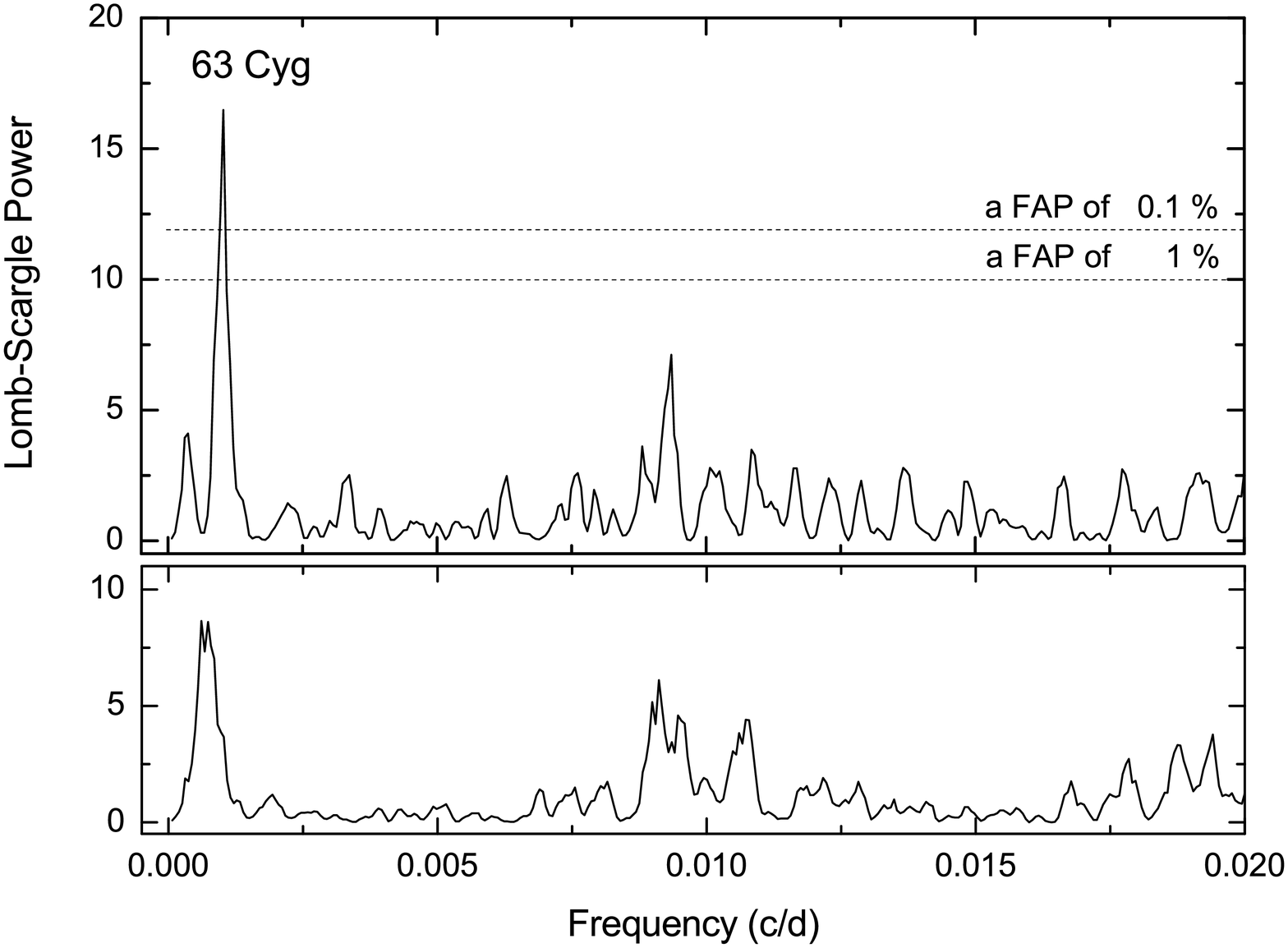}
      \caption{Lomb-Scargle periodogram of the RV measurements for 63 Cyg. The periodogram shows significant power at a period of 982.8 days (\emph{top panel}) and after subtracting the main frequency variations (\emph{bottom panel}). The horizontal lines indicate FAP thresholds of 0.1 and 1\%.
      }
         \label{power2}
   \end{figure}
%
%
\begin{table}
\begin{center}
\caption{RV measurements for 63 Cyg from October 2004 to May 2014 using the BOES.}
\label{tab3}
\begin{tabular}{rrrrrr}
\hline\hline

 JD         & $\Delta$RV  & $\pm \sigma$ &        JD   & $\Delta$RV  & $\pm \sigma$  \\
 $-$2 450 000 & m\,s$^{-1}$ &  m\,s$^{-1}$ & $-$2 450 000  & m\,s$^{-1}$ &  m\,s$^{-1}$  \\
\hline
3301.0456 &   176.4 &     4.1 &  5671.3300  &    367.9  &    9.0  \\
3301.0520 &   168.6 &     3.9 &  5841.9930  &   $-$367.7  &    4.5  \\
3331.9292 &   299.1 &     5.0 &  5932.9265  &   $-$607.9  &    5.8  \\
3331.9365 &   297.9 &     5.2 &  6177.0317  &    107.6  &    5.0  \\
3896.2348 &  $-$336.1 &     5.5 &  6261.0609  &    239.3  &   14.5  \\
3899.3163 &  $-$314.8 &     5.6 &  6334.3845  &    269.3  &    9.2  \\
4027.0155 &  $-$553.2 &     4.6 &  6343.3885  &    229.1  &   13.0  \\
4263.1944 &   287.8 &     7.8 &  6347.3861  &    345.7  &    9.8  \\
4751.0915 &  $-$447.9 &     8.6 &  6347.3916  &    347.5  &    8.8  \\
4848.9094 &  $-$670.2 &     6.8 &  6616.9001  &     18.6  &    6.5  \\
4930.3505 &  $-$592.8 &     5.9 &  6616.9035  &     17.1  &    7.2  \\
4971.1938 &  $-$558.5 &     5.9 &  6678.9125  &    $-$16.0  &   15.6  \\
4995.1952 &  $-$575.9 &     5.7 &  6680.9067  &     $-$7.6  &    9.5  \\
5131.0074 &   $-$58.1 &     4.1 &  6680.9131  &     $-$8.2  &    8.3  \\
5356.2036 &   682.9 &     5.5 &  6680.9170  &     $-$8.8  &    8.4  \\
5455.2002 &  1046.0 &     4.7 &  6790.2199  &   $-$301.4  &   12.6  \\
5553.9435 &   812.1 &     8.4 &  6790.2252  &   $-$288.5  &   17.9  \\

\hline

\end{tabular}
\end{center}
\end{table}
%

%
   \begin{figure}
   \centering
   \includegraphics[width=8cm]{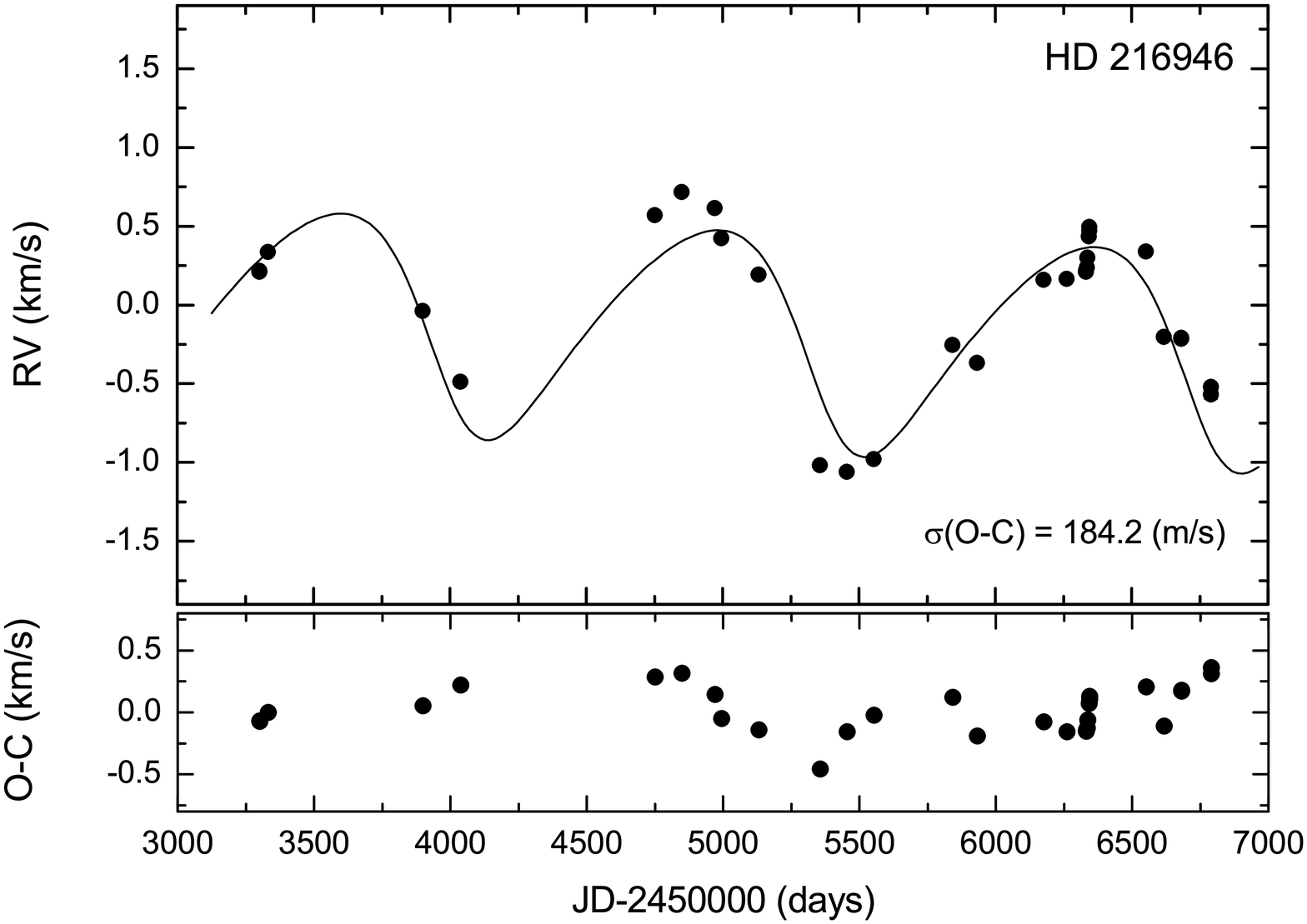}
      \caption{Variations of the RV curve (\emph{top panel}) and rms scatter of the residual (\emph{bottom panel}) for HD 216946 from October 2004 to May 2014. The solid line is the orbital solution with a period of 1382.0 days and an eccentricity of 0.22.
              }
         \label{rv3}
   \end{figure}
%
%
   \begin{figure}
   \centering
   \includegraphics[width=8cm]{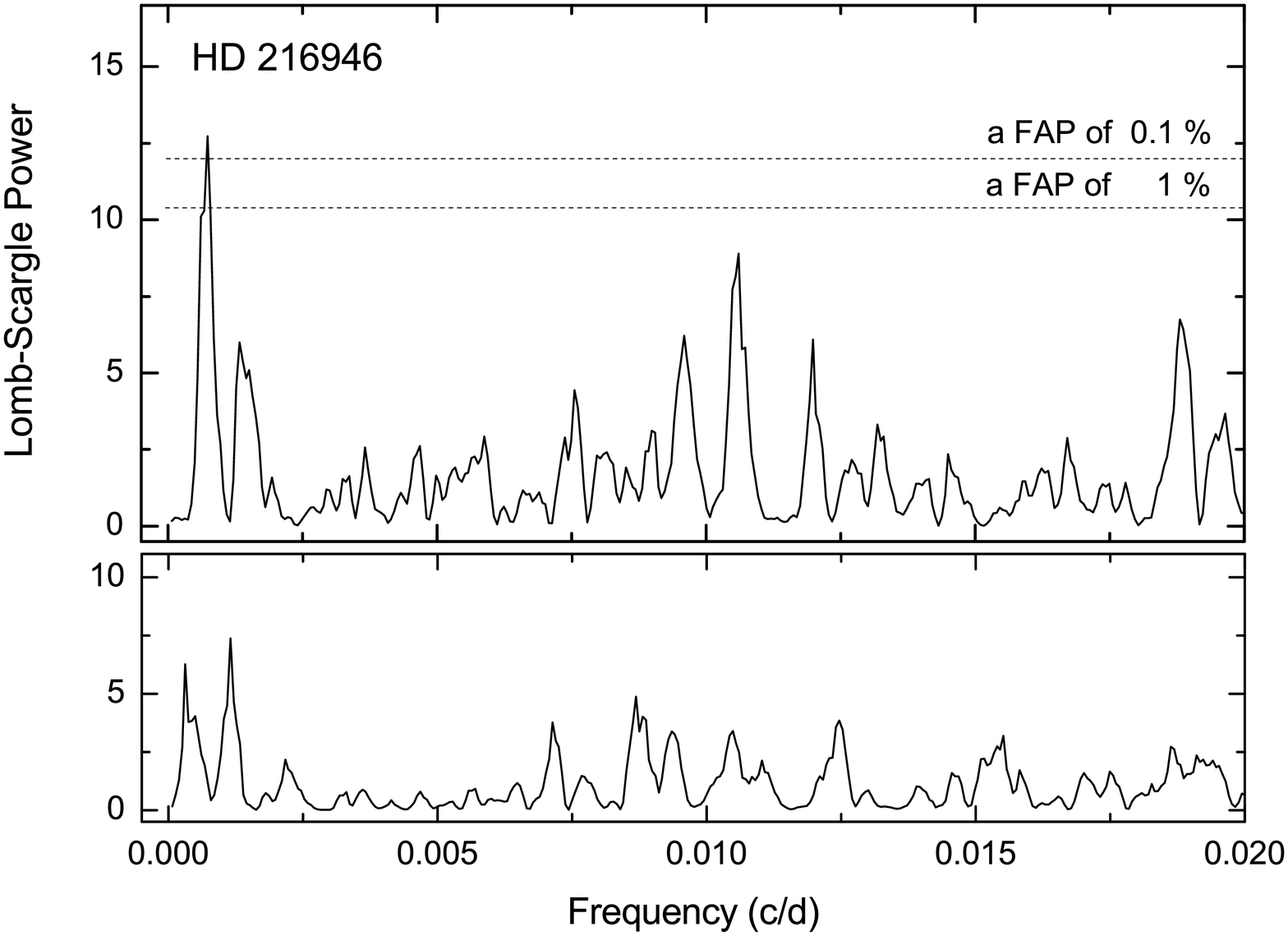}
      \caption{Lomb-Scargle periodogram of the RV measurements for HD 216946. The periodogram shows significant power at a period of 1382.0 days (\emph{top panel}) and after subtracting the main frequency variations (\emph{bottom panel}). The horizontal lines indicate FAP thresholds of 0.1 and 1\%.
      }
         \label{power3}
   \end{figure}
%

The Lomb-Scargle periodogram (Lomb 1976; Scargle 1982) is a useful tool for  investigating periodic variations in  unequally spaced time-series data as well as to evaluate their significance. The Lomb-Scargle periodogram of the RV measurements for HD 3574 shows  significant power near the frequency of 0.000943 c d$^{-1}$ (Fig.~\ref{power1}). This peak corresponds to a period of 1061.2 $\pm$ 5.4 days.
We also estimated a false-alarm probability (FAP) of the detected periods using a bootstrap randomization procedure (K{\"u}rster et al. 1999). We shuffled the RV values 200~000 times keeping the time fixed and noted the number of the Lomb-Scargle periodogram for the random data that had power higher than the real data. This gives a better estimate of a FAP of $<$ 3 $\times$ 10$^{-2}$\%.
An orbital fit to the RV measurements yield a $K$ amplitude of 376.0 $\pm$ 14.3 m s$^{-1}$  and an eccentricity, $e$ = 0.32 $\pm$ 0.06.
Assuming a stellar mass of 7.7 $\pm$ 0.4 $M_{\odot}$ (Tetzlaff et al. 2011), we derived a minimum mass of a substellar companion $m$~sin~$i$ = 69.7 $\pm$ 6.6 $\it M_\mathrm{Jup}$ at an orbital distance $a$ = 4.0 $\pm$ 0.1 AU from the host star.
The largest uncertainty is from the unknown orbital inclination (\emph{i}). If we assume that the orbit is randomly oriented, there is about a 51\% chance that the true mass exceeds 80 $\it M_\mathrm{Jup}$ (\emph{i} $\leq$ 61), the border between brown dwarf and stellar mass regimes.
The RV measurements for HD~3574 including the orbital fit are plotted in Fig.~\ref{rv1}.


The RV variations of 63 Cyg can be well fit by a Keplerian orbit with $P$ = 982.8 $\pm$ 6.9 days ($f_{1}$ = 0.001018 c~d$^{-1}$), $K$ = 742.0 $\pm$ 54.6 m s$^{-1}$, and $e$ = 0.29 $\pm$ 0.12. It has an FAP lower than 10$^{-3}$\% and there was no instance where the random data had higher power. The RV variations and the Lomb-Scargle periodogram for 63 Cyg are shown in Figs.~\ref{rv2} and ~\ref{power2}.


The time series of our RV measurements for HD 216946 are listed in Table~\ref{tab4} and are shown in Fig.~\ref{rv3}. The Lomb-Scargle periodogram of the data (Fig.~\ref{power3}) exhibits a dominant peak corresponding to a period of 1382.0 $\pm$ 12.2 days ($f_{1}$ = 0.000723 c~d$^{-1}$) and an FAP of $\sim$ 5 $\times$ 10$^{-2}$\%. An orbital solution  results in $K$ = 699.5 $\pm$ 50.3 m s$^{-1}$ and $e$ = 0.22 $\pm$ 0.12. Assuming a stellar mass of 6.8 $\pm$ 1.0 $M_{\odot}$  (Tetzlaff et al. 2011) results in a companion minimum mass of $m$ sin $i$ = 134.1 $\pm$ 27.5 $\it M_\mathrm{Jup}$ and a semi-major axis of 4.6 $\pm$ 0.2 AU.

We note that there are large scatters of 110.1, 169.7, and 184.2 m s$^{-1}$ in the residuals after subtracting the dominant period from each RV data set. Periodogram analyses reveal no additional significant periods in the data. Large intrinsic RV scatter is common for evolved stars because of stellar oscillations. We estimated the amplitude and period of stellar oscillations in these stars using the scaling relations of Kjeldsen \& Bedding (2011): 170 and 10 (HD 3574), 137 and 8 (63 Cyg), and 350 m\,s$^{-1}$ and 23 days (HD 216946), respectively.
These results demonstrate that the high rms scatter about the orbital solution can easily be accounted for entirely by stellar oscillations.
These increase toward later spectral types and evolved stars (Hatzes et al. 2005; Hekker et al. 2006; Lee et al. 2013).

%
\begin{table}
\begin{center}
\caption{RV measurements for HD 216946 from October 2004 to May 2014 using the BOES.}
\label{tab4}
\begin{tabular}{rrrrrr}
\hline\hline

 JD         & $\Delta$RV  & $\pm \sigma$ &        JD & $\Delta$RV  & $\pm \sigma$  \\
 $-$2 450 000 & m\,s$^{-1}$ &  m\,s$^{-1}$ & $-$2 450 000  & m\,s$^{-1}$ &  m\,s$^{-1}$  \\
\hline
3301.0667  &    217.3 &       6.1 &    6261.0731  &    166.3  &      9.9 \\
3301.0735  &    211.0 &       5.3 &    6331.9302  &    216.6  &      9.6 \\
3333.0089  &    335.6 &       5.2 &    6332.9187  &    211.2  &      9.6 \\
3899.2855  &    $-$36.9 &       6.7 &    6334.9237  &    239.2  &     12.1 \\
4038.1784  &   $-$488.0 &       6.5 &    6336.9071  &    302.1  &      9.3 \\
4751.1239  &    569.5 &      10.5 &    6342.9085  &    436.8  &      8.8 \\
4848.9428  &    717.7 &       7.2 &    6343.9172  &    472.6  &      9.6 \\
4971.2603  &    616.5 &       7.5 &    6344.9193  &    496.2  &     10.9 \\
4995.2854  &    422.9 &       6.5 &    6551.9765  &    340.2  &      9.2 \\
5131.0296  &    193.5 &       6.4 &    6616.9139  &   $-$203.3  &      9.2 \\
5356.2355  &  $-$1017.1 &       7.8 &    6616.9167  &   $-$202.0  &      9.3 \\
5455.2171  &  $-$1060.0 &       7.5 &    6680.9216  &   $-$210.3  &     10.6 \\
5553.9828  &   $-$976.3 &       8.5 &    6680.9236  &   $-$214.8  &      8.9 \\
5842.0035  &   $-$251.6 &       6.3 &    6680.9255  &   $-$209.2  &     10.2 \\
5932.9371  &   $-$366.8 &       9.3 &    6790.2932  &   $-$567.7  &     29.6 \\
6177.0440  &    158.8 &       6.7 &    6790.2988  &   $-$518.6  &     23.4 \\
\hline

\end{tabular}
\end{center}
\end{table}
%

%
%
%
%


\section{Discussion}
From our  RV survey of evolved stars obtained with BOES over approximately ten years, we find that HD~3574, 63~Cyg, and HD~216946 show evidence of low-amplitude (several hundreds of m s$^{-1}$) and long-period (about a thousand days) RV variations. The RV variations in evolved stars can be caused by stellar pulsations, rotational modulations by inhomogeneous surface features, or substellar-mass companions.

Very long-period and low-amplitude RV variations are common among evolved stars. It is unlikely that these are caused by radial pulsations because the estimated period of the fundamental radial mode is several days. However, we cannot exclude the possibility of non-radial modes, or other forms of exotic, and heretofore unknown pulsation modes.  Lee et al. (2008; 2012) speculated the existence of a yet unrecognized mechanism that drives low-amplitude and long-period variations of the radius in Polaris and $\alpha$ Per. Clearly, more detailed studies are needed to understand possible oscillations in these stars.


%
RV modulations in giant stars are often caused by spots instead of by orbiting planets (Bonfils et al. 2011). Hence, it is most important to measure activity indicators for these evolved stars. To do this several diagnostics need to be applied (i.e., line bisector variations and photometric variations) to distinguish between stellar activity and RV variations that are caused by orbital motion.
However, line bisector analysis is generally ineffective for evolved stars with slow rotational velocities therefore photometric data are more reliable. Unfortunately, we have no contemporaneous photometric measurements, only \emph{HIPPARCOS}
data from several years earlier.
More research may help in understanding the mechanisms of the variations of the giant HD 3574 and supergiant 63 Cyg.

%

%
The M supergiant HD 216946 reveals different long-term variations in spectroscopic and photometric measurements. We considered two interpretations for these variations.
First, HD~216946 was discovered to be variable in the optical band: long-period variations of $\sim$ 1100 days in the \emph{HIPPARCOS} and 1601 days in the APT-80 data from Messina (2007).
These variations are similar to that of a semi-regular variable in late-type supergiants (SRC) or a long secondary-period variable (LSPV). SRC are M supergiants that show semiregular, multiperiodic photometric variations ranging from several ten to several thousand days (i.e., BC Cyg, $\alpha$ Ori, $\mu$ Cep).
Many are surrounded by dense circumstellar dust shells created by their ongoing mass loss, and the variables are also believed to be the precursors of Type II supernovae (Chevalier 1981).
However, unlike the one-magnitude variation in V band usually observed in the SRC variable, HD~216946  just shows 0.05$-$0.08 mag variations in the photometric results. It is unclear whether SRC pulsations are the cause of the observed RV variations.
On the other hand, Kiss et al. (2006) showed that  $\sim$ 25\% of pulsating red supergiants show periodic brightness changes characterized by two distinct time scales of a few hundred days (first-period mode) and longer than about 1000 days (secondary-period mode), known as LSPV. For HD 216946, the first period is related to pulsational modes and the long secondary period may be contributed by a near-UV excess (Messina 2007). This means that HD~216946 is closer to the LSPV than to the SRC terms of both  photometric and RV variability. Since HD 216946 is known to be a multiperiodic variable, it may be that the RV period is just another pulsation mode that was not present during the photometric measurements.

Second, the low-amplitude and long-period RV variations for HD~216946 may result from the rotational modulation of surface activities such as large, cool spots.
The nature of spots on evolved stars or their lifetimes are not known. Spots on very active stars can be long-lived.
Hatzes (1995) suggested that the spot on V410 Tau has survived on the stellar surface for $\sim$ 20 yr based on a comparison of Doppler images. A similarly long-lived spot can be excluded for HD 216946 since the 1382\,d period would have been detected in the earlier photometric measurements. Messina (2007) concluded that HD 216946 cannot be ascribed to the rotational modulation of cool spots by using the model variations of UBV magnitudes vs. wavelength. Generally, cool spots cause an amplitude increase towards shorter wavelengths because of the higher contrast between spots and photosphere, which is clearly different from HD 216946. Currently, the most likely explanations
for the RV variations for HD 216946 are a mode of long-period stellar pulsations or a stellar companion. Only more RV measurements for this star, preferably with contemporaneous photometric measurements, can distinguish between these two hypotheses.


\begin{acknowledgements}
      BCL acknowledges partial support by the KASI (Korea Astronomy and Space Science Institute) grant 2013-9-400-00. Support for MGP was provided by the National Research Foundation of Korea to the Center for Galaxy Evolution Research (No. 2010-0027910).
      This research made use of the SIMBAD database, operated at the CDS, Strasbourg, France.

\end{acknowledgements}
%


\end{document}